\documentclass[10pt,singlespace,twocolumn]{article}

\usepackage{graphicx}
\usepackage{amssymb}
\usepackage{enumerate}
\usepackage{color}

\newcommand{\be}{\begin{equation}}
\newcommand{\ee}{\end{equation}}
\newcommand{\n}{\noindent}
\newcommand{\dd}{{\rm d}}

\begin{document}
\title{Morphology and Interaction between Lipid Domains}         
\author{\small Tristan S. Ursell$^1$,\,\,\,\,William S. Klug$^2$\,\,\,\,and Rob Phillips$^1$\footnote{Address correspondence to: phillips@pboc.caltech.edu}\\ 
\small $^1$Department of Applied Physics, California Institute of Technology, Pasadena, CA 91125\\
\small $^2$Department of Mechanical and Aerospace Engineering, Program in Biomedical Engineering, \\ \small and California NanoSystems Institute, University of California Los Angeles, Los Angeles, CA 90095}
\date{}
\maketitle

{\footnotesize Cellular membranes are a heterogeneous mix of lipids, proteins and small molecules.  Special groupings of saturated lipids and cholesterol form a liquid-ordered phase, known as `lipid rafts,' serving as platforms for signaling, trafficking and material transport throughout the secretory pathway.  Questions remain as to how the cell maintains heterogeneity of a fluid membrane with multiple phases, through time, on a length-scale consistent with the fact that no large-scale phase separation is observed. We have utilized a combination of mechanical modeling and {\it in vitro} experiments to show that membrane morphology can be a key player in maintaining this heterogeneity and organizing such domains in the membrane.  We demonstrate that lipid domains can adopt a flat or dimpled morphology, where the latter facilitates a repulsive interaction that slows coalescence and tends to organize domains. These forces, that depend on domain morphology, play an important role in regulating lipid domain size and in the lateral organization of lipids in the membrane.}\\


\small{
The plasma and organelle membranes of cells are composed of a host of different lipids, lipophilic molecules and membrane proteins \cite{Singer1972}.  Together, they form a heterogeneous layer capable of regulating the flow of materials and signals into and out of the cell.  Lipid structure and sterol content play a key role in membrane organization, where steric interactions and energetically costly mismatch in the hydrophobic structure of lipid tails result in lateral phase-separation.   Saturated lipids and cholesterol are sequestered into liquid-ordered ($L_{\mbox{\footnotesize o}}$) domains, often known as `lipid rafts', from an unsaturated liquid-disordered ($L_{\mbox{\footnotesize d}}$) phase \cite{Veatch2003,Bacia2005,Baumgart2003}.   Domains composed of saturated sphingolipids and cholesterol, with sizes in the range of $\sim50-500\,\mbox{nm}$, have been implicated in a range of biological processes from lateral protein organization and virus uptake to signaling and plasma-membrane tension regulation \cite{Simons1997,SensPRE2006,Sheetz1999,Simons2004,Schlegel1998,Sprong2004,Chazal2003,RaoTraffic2004,Dietrich2001,Park1998,Helms2004,Lucero2004,Gaus2003}.  How the cell maintains the lateral heterogeneity of lipids over time, and what physical mechanism might be responsible for the spatial organization of these domains, challenges classical theories of phase-separation and `domain ripening' (such as Cahn-Hilliard kinetics \cite{Bray2002}).  The maintenance of lateral heterogeneity is thought to arise from a combination of lipid recycling and energetic barriers to domain coalescence \cite{SensPRL2005,Gheber1999,Dietrich2002} (potentially provided by transmembrane proteins \cite{KusumiBiophysJ2004}), resulting in a stable distribution of domain sizes.  The precise origin of this energy barrier and the nature of its dependence on membrane elastic properties remains unclear. 

The simplest physical model that describes the evolution of lipid domain size and position predicts that domains diffuse and coalesce, such that the number of domains constantly decreases, while the average domain size constantly increases \cite{Bray2002}.  Indeed, models of two-dimensional phase separation have been studied in detail for many physical systems \cite{SaguiPRL1995,LaradjiPRL2004,Seul1995,SaguiPRE1994}, and where the phase boundary is unfavorable and characterized by an energy per unit length \cite{Kuzmin2005}, the domain size grows continuously ($\propto t^{1/3}$) \cite{Bray2002,SeulPRL1994,Foret2005}.  However, membranes can adopt three-dimensional morphologies that affect the kinetics of phase separation \cite{Harden2005,Taniguchi1996,Gompper2001,Riegada2005,Laradji2006}.  In those cases where morphology is considered as part of the phase separation model, novel coalescence kinetics emerge \cite{Taniguchi1996}.  Experimentally, model membranes have shown that nearly complete phase separation on the surface of a cell-sized vesicle can be reached in as little as one minute \cite{Veatch2003}.  This seems inconsistent with the fact that on the cell surface, much smaller domains persist on that same time-scale \cite{Dietrich2002} and no large-scale phase separation is observed.   With these facts in mind, our central questions are:  how can membranes that have phase-separated maintain their lateral heterogeneity on long time scales and short length scales?   Are there membrane-mediated ({\it i.e.} elastic) forces that inhibit coalescence and spatially organize domains?

We begin to answer these questions by examining the energetics of the membrane using a linear elastic model. A phase-separated membrane is endowed with bending stiffness, membrane tension, an energetic cost at the phase boundary, and domains of a particular size.  Membrane bending and tension establish a natural length-scale over which a morphological instability develops that switches domains from a flat to `dimpled' shape, similar to classical Euler buckling \cite{Freund} (see Figure \ref{fig1}).  The dimpling instability is size-selective and `turns on' a membrane-mediated interaction that inhibits domain coalescence.  This transition is a precursor to budding, and is distinct from transitions that require spontaneous curvature. While variations in membrane composition may change specific parameter values, the mechanical effects we describe are generic.  Thus, these systems exhibit shape-dependent coarsening kinetics, that are relevant for a broad class of two-dimensional binary fluid systems. The interaction between domains is a mechanical effect, and we use a model treating dimpled domains as curved rigid inclusions to distill the main principles governing this interaction.  The confluence of membrane properties required for this morphological change and its attendant forces lies squarely in the biological regime.  Experimentally, we use a model mixture of lipids and cholesterol to show that such an interaction exists between dimpled domains and is well approximated by a simple model. Combined with lipid recycling \cite{SensPRL2005}, we offer elastic interactions as a mechanism for the maintenance of lipid lateral-heterogeneity and organization of domains in cellular membranes.

\begin{figure}
\begin{center}
\includegraphics[width=3.6in]{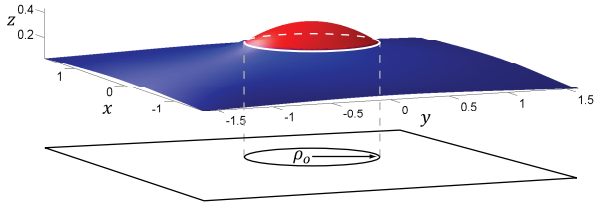}
\caption{\footnotesize Three dimensional rendering of a dimpled lipid domain in dimensionless coordinates.  For a domain (shown in red), a competition between bending, membrane tension, and phase boundary line tension results in a morphological transition from a flat to a dimpled state as depicted above.  The dimple costs bending energy but relieves line tension by reducing the phase boundary length (shown as a white line around the domain). This morphology facilitates interactions between domains that significantly alter the kinetics of coalescence and lateral lipid organization.  The projected domain radius is $\rho_o=r_o/\lambda_2$.}
\label{fig1}
\end{center}
\end{figure}

The first section of the paper outlines the energetic contributions to the mechanical model, and predicts the conditions under which domain dimpling occurs.  The second section outlines how dimpled domains facilitate an elastic interaction and compares the model interaction to our measurements made in phase-separated giant unilamellar vesicles.\\

\n
{\bf The Elastic Model and Morphological Transitions\\} 
The energetics of a lipid domain are dominated by a competition --- on one hand the applied membrane tension and bending stiffness both energetically favor a flat domain; on the other hand the phase boundary line tension prefers any domain morphology (in 3D) that reduces the boundary length.  We use a continuum mechanical model that couples these effects, relating the energetics of membrane deformation to domain morphology.  As we will show, this competition results in a morphological transition from a flat to dimpled domain shape, where two dimpled domains are then capable of interacting elastically. 

Lipid domains in a liquid state naturally adopt a circular shape to minimize the phase boundary length \cite{Veatch2003}, allowing us to formulate our continuum mechanical model in polar coordinates.  We employ a {\it Monge} representation, where the membrane mid-plane is described by a height function $h({\bf r})$ in the limit of small membrane deformations ({\it i.e.} $|\nabla h|<1$).  With this height function, we characterize how membrane tension, bending, spontaneous curvature and line tension all contribute to domain energetics.  

Changes in membrane height alter the projected area of the membrane and hence do work against the applied membrane tension, resulting in an increase in energy written as
\be{
G_{\mbox{\tiny tens}}=\pi\tau\left(\int_0^{r_o}(\nabla h_1)^2r\dd r+\int_{r_o}^{\infty}(\nabla h_2)^2r\dd r\right),
}\ee
where $\tau$ is the constant membrane tension, $r_o$ is the projected radius of the domain, $h_1$ is the height function of the domain and $h_2$ is the height function of the surrounding membrane \cite{Boal,Wiggins2005}. Membrane curvature is penalized by the bending stiffness with a bending energy written as \cite{Helfrich1973,Boal}
\be{
G_{\mbox{\tiny bend}}=\pi\kappa_b^{\mbox{\tiny (2)}}\left(\sigma\int_0^{r_o}\left(\nabla^2 h_1\right)^2r\dd r +\int_{r_o}^{\infty}\left(\nabla^2 h_2\right)^2r\dd r\right).
}\ee
Our model allows the domain and surrounding membrane to have differing stiffnesses, $\kappa_b^{\mbox{\tiny (1)}}$ and $\kappa_b^{\mbox{\tiny (2)}}$ respectively, characterized by the parameter $\sigma=\kappa_b^{\mbox{\tiny (1)}}/\kappa_b^{\mbox{\tiny (2)}}$, and from this point on we drop the superscript on $\kappa_b^{\mbox{\tiny (2)}}$. Recent experiments suggest that the bending moduli of a cholesterol-rich domain and the surrounding membrane are roughly equal \cite{Baumgart2003,Groves2006}, and hence for simplicity, we assume the bending moduli of the two regions are equal ({\it i.e.} $\sigma=1$), unless otherwise noted.  In addition to bending stiffness, the domain may exhibit a preferred `spontaneous' curvature due to lipid asymmetry or protein binding \cite{Farsad2003,Laradji2006}.  The contribution of domain spontaneous curvature can be simplified to a boundary integral, which couples to the overall curvature field by
\be{
G_{\mbox{\tiny spont}}= -2\pi\sigma\kappa_bc_o\int_0^{r_o}\left(\nabla^2h_1\right) r\dd r=-2\pi\sigma\kappa_bc_or_o\epsilon,
}\ee
where $c_o$ is the spontaneous curvature of the domain and $\epsilon$ is the membrane slope at the phase boundary as shown in Figure \ref{fig1}.  Further, we assume the saddle-splay curvature moduli are equal in the two regions, yielding no dependence on Gaussian curvature.  In principle, this contribution could be accounted for with a boundary term, explored in detail in the supplementary information (SI).  The phase boundary line tension is applied to the projected circumference of the domain, as shown in Figure \ref{fig1}, by $G_{\mbox{\tiny line}}=2\pi r_o\gamma$ where $\gamma$ is the energy per unit length at the phase boundary.  

Finally, a constraint must be imposed that relates the actual domain area, $\mathcal{A}$, to the projected domain radius $r_o$. The energetic cost to change the area per lipid molecule is high ($\sim50-100\,k_BT/\mbox{nm}^2$ where $k_B=1.38\times10^{-23}\,J/\mbox{K}$ and $T=300\,\mbox{K}$ \cite{Evans2000}), hence we assume the domain area is conserved during any morphological change (see SI for details).  We impose this constraint using a Lagrange multiplier, $\tau_o$, with units of tension by
\be{
G_{\mbox{\tiny area}}=\tau_o\left(\pi\int_0^{r_o}(\nabla h_1)^2r\dd r+\pi r_o^2-\mathcal{A}\right).
}\ee
This results in an effective membrane tension within the domain $\tau_1=\tau+\tau_o$, which must be negative to induce dimpling.  Examining the interplay between bending and membrane tension, we see that two natural length scales emerge - within the domain we define $\lambda_1=\sqrt{\sigma\kappa_b/\tau_1}$ and outside the domain we define $\lambda_2=\sqrt{\kappa_b/\tau}$.  These length scales allow us to define the relevant dimensionless parameters in this system. 

The total free energy of an elastic domain and its surrounding membrane is then the sum of these five terms, $G=G_{\mbox{\tiny tens}}+G_{\mbox{\tiny bend}}+G_{\mbox{\tiny spont}}+G_{\mbox{\tiny line}}+G_{\mbox{\tiny area}}$.  Details on all the terms in the free energy can be found in the SI.   With this free energy in hand, we examine how the morphology of a circular domain evolves as we tune domain size and the elastic properties of the membrane. 

The height field and radius can be rescaled by the elastic decay lengths such that the Euler-Lagrange equation for the domain can be written in the parameter-free form $\nabla^2(\nabla^2+\beta^2)\eta_1=0$, while the equation for the surrounding membrane is $\nabla^2(\nabla^2-1)\eta_2=0$, where the dimensionless variables are defined by $\lambda_2\eta_i=h_i$, $\lambda_2\rho=r$, $\lambda_2\rho_o=r_o$ and $\beta=i\lambda_2/\lambda_1$. Using the same dimensionless notation, the energy from line tension and spontaneous curvature can be written as $G_{\mbox{\tiny line}}=2\pi\kappa_b\rho_o\chi$ and $G_{\mbox{\tiny spont}}=-2\pi\sigma\kappa_b\epsilon\rho_o\upsilon_o$, with $\upsilon_o=\lambda_2 c_o$ and $\chi=\gamma\lambda_2/\kappa_b$.  The dimensionless line tension, $\chi$, is simply a rescaled version of the line tension $\gamma$ and is one of two key parameters that characterize the morphological transition; the dimensionless domain area, $\alpha=\mathcal{A}/\lambda_2^2$, is the second key parameter.

The admissible solutions for $\eta_1(\rho)$ and $\eta_2(\rho)$ are zeroth order Bessel functions $J_0(\beta\rho)$ and $K_0(\rho)$, respectively, with the boundary conditions $|\nabla \eta_1(0)| = |\nabla \eta_2(\infty)| = 0$ and $|\nabla \eta_1(\rho_o)| = |\nabla \eta_2(\rho_o)| =\epsilon$.  The boundary slope, $\epsilon$, is the parameter that indicates the morphology of the domain; $\epsilon=0$ indicates a flat domain, while $0<|\epsilon|\lesssim 1$ indicates a dimpled domain.  The five contributions to membrane deformation energy yield a relatively simple expression for the total free energy, given by
\begin{eqnarray}
\label{freeentot}
G&=&\pi\kappa_b\rho_o\left[\epsilon^2\left(\sigma\beta\frac{J_0(\beta\rho_o)}{J_1(\beta\rho_o)}+\frac{K_0(\rho_o)}{K_1(\rho_o)}\right)+2(\chi-\epsilon\sigma\upsilon_o)\right]\\
&&-\kappa_b(\sigma\beta^2+1)(\pi\rho_o^2-\alpha)\nonumber.
\end{eqnarray}
Mechanical equilibrium is enforced by rendering the energy stationary with respect to unknown parameters $\epsilon$, $\rho_o$, and $\beta$,  
\be{
\label{eqsys}
\frac{\partial G}{\partial\epsilon}=0,  \quad \frac{\partial G}{\partial\rho_o}=0, \quad \frac{\partial G}{\partial\beta}=0.
}\ee
These equilibrium equations physically correspond to torque balance at the phase boundary, lateral force balance at the phase boundary and domain area conservation, respectively.

Analysis of the equilibrium equations reveals a {\it second-order} transition at a critical line-tension, $\chi_c$, as shown in Figure \ref{fig2}.  For $\chi$ less than this critical value, only the flat, trivial solution with $\epsilon=0$ exists.  At $\chi_c$ a non-trivial solution describing buckled or dimpled morphologies emerges.  For zero spontaneous curvature, the bifurcation is defined by a transcendental characteristic equation
\be{
\sigma\beta\frac{J_0(\beta\rho_o)}{J_1(\beta\rho_o)}+\frac{K_0(\rho_o)}{K_1(\rho_o)}=0 ,
\label{phaseeq}
}\ee
with $\beta=\sqrt{(\chi_c/\rho_o-1)/\sigma}$ and $\rho_o=\sqrt{\alpha/\pi}$.  For a given dimensionless domain area, $\alpha$, this defines the critical line tension required to dimple the domain. In Figure \ref{fig2}a(inset), this relation is used to generate a morphological phase diagram that shows where in the space of dimensionless domain area and line tension we find the discontinuous transition ({\it i.e.} bifurcation) from a flat domain, to a dimpled domain.  Near the morphological transition the boundary slope grows as $|\epsilon|\propto\sqrt{\chi/\chi_c-1}$, indicating that a dimple rapidly deviates from the flat state.  The transition is symmetric, in that both possible dimple curvatures have the same energy, and hence the domain is equally likely to dimple upwards or downwards.  In the experimentally relevant limit of small dimensionless domain area, the complexity of eqn.~\ref{phaseeq} is reduced to
\be{
\label{chicrit}
\chi_c\sqrt{\alpha}=\frac{\gamma_c}{\kappa_b}\sqrt{\mathcal{A}}\simeq8\sigma\sqrt{\pi}.
}\ee
This leads to the conclusion that the dominant parameter governing domain dimpling at zero spontaneous curvature is $\chi\sqrt{\alpha}$. For a small domain, the dimpling transition is directly regulated by domain area, the bending modulus, and line tension, but only weakly depends on applied membrane tension. Intuitively, domains dimple when line tension or domain size increase (subject to small $\alpha$), as shown in Figure \ref{fig2}a(inset).  Likewise, a decrease in bending stiffness, due, for instance, to changes in membrane sterol content \cite{Simons2000,Lange2004}, can also induce dimpling. The effects of applied membrane tension are weak because the change in projected area upon dimpling does not lead to a significant energy cost relative to the cost of bending and line tension.

If membrane elastic properties are fixed ({\it i.e.} fixed $\kappa_b$, $\tau$ and $\gamma$), the dimpling-induced interactions `turn on' only after a critical domain size is achieved.  This scenario is encountered when two domains, too small to dimple on their own, diffusively coalesce into a larger domain capable of dimpling and hence interacting. Indeed, such a size-selective coalescence mechanism was observed recently in model membrane vesicles \cite{Yanagisawa2007}.  This constitutes a distinct class of coarsening dynamics, where classical diffusion-limited kinetics are obeyed until the domain size distribution has matured past the critical size for dimpling - then domain coalescence is a relatively slow, interaction-limited process.

\begin{figure}
\begin{center}
\includegraphics[width=3in]{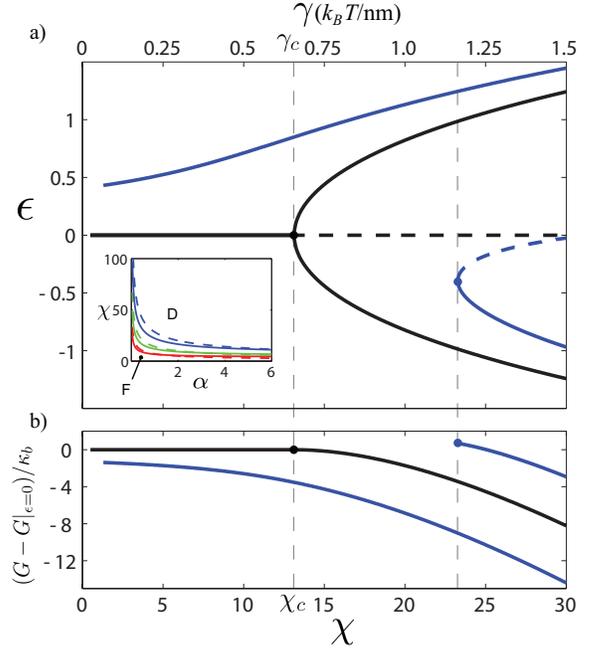}
\caption{\footnotesize Bifurcation diagram for dimpling transition at constant area ($\alpha=\pi/4$, $\kappa_b=25\,k_BT$, $\lambda_2=500\,\mbox{nm}$, $\sigma=1$).  Constant line tension and increasing area produces a qualitatively similar graph. a) At zero spontaneous curvature ($\upsilon_o=0\rightarrow\mbox{black}$) the bifurcation is symmetric, the upper and lower branches are at the same energy, and $\epsilon=0$ becomes unstable above the critical point (horizontal black dashed line).  With finite spontaneous curvature ($\upsilon_o=2$, $c_o=(250\,\mbox{nm})^{-1}\rightarrow\mbox{blue}$) the lower energy branch (upper) has non-zero $\epsilon$ for all line tensions, asymptoting to the $\upsilon_o=0$ branch.  At a line tension slightly higher than $\chi_c$ for the $\upsilon_o=0$ case, a bifurcation yields a higher energy dimple with the opposite curvature as $\upsilon_o$ (indicated by the second vertical dashed line). Inset:  Equilibrium phase diagrams for $\sigma=0.5$(red), $\sigma=1$(green), and $\sigma=2$(blue) (the dashed lines are the approximation of eqn.~\ref{chicrit}) showing flat (F) and dimpled (D) domains. b) Energy difference between the flat and dimpled state, normalized by $\kappa_b$, for domains with and without spontaneous curvature ($\upsilon_o=0\rightarrow\mbox{black}$; $\upsilon_o=2\rightarrow\mbox{blue}$).}
\label{fig2}
\end{center}
\end{figure}

For the model domain considered in Figure \ref{fig2}, with area $\alpha=\pi/4$ ($r_o\simeq250\,\mbox{nm}$), the critical dimensionless line tension is $\chi_c\simeq13$, corresponding to a critical line tension of $\gamma_c\simeq0.65\,k_BT/\mbox{nm}$.  This value compares well with both theoretical estimates of the line tension \cite{Kuzmin2005,Lipowsky1992}, and the higher side of experimentally measured values \cite{Baumgart2003,Tian2007,Schwille2007}. 

Spontaneous curvature does not affect the Euler-Lagrange equations, and hence will not effect the class of equilibrium membrane shapes. However, domains with zero and nonzero spontaneous curvature exhibit qualitatively different behavior.  Biological membranes can be asymmetric with respect to leaflet composition \cite{Simons2000,Simons1997,WangBJ2007}, endowing a domain with potentially large spontaneous curvature. The energetic contribution from spontaneous curvature takes the form of an additional line tension depending linearly on the slope taken by the domain boundary, $\epsilon$. This breaks the symmetry of the membrane, giving an energetic preference to a dimple with the same curvature as the spontaneous curvature, and eliminating the trivial $\epsilon=0$ solution even at small line-tensions.  As line tension increases, a bifurcation produces a second, stable, higher-energy dimple of the opposite curvature as $\upsilon_o$.  The more energetically stable branch of this transition corresponds to a dimpled state for {\it all} values of line tension and non-zero values of domain area, as demonstrated in Figure \ref{fig2}a.  This predicts that as soon as a domain with finite spontaneous curvature forms, it dimples, regardless of size, and begins to experience interactions with any nearby dimpled domains.  It is reasonable to expect that domains with similar composition will have similar spontaneous curvature, and hence form dimples whose curvature has the same sign.  As we will show, dimples whose curvature has the same sign tend to interact repulsively.  Such a mechanism of coalescence inhibition was observed recently in simulation \cite{Laradji2006}.

This indicates that control of spontaneous curvature via domain composition or protein binding can regulate dimpling and hence domain interaction \cite{McMahon2005,WangBJ2007}.   Indeed, recent experimental \cite{Baumgart2007} and theoretical \cite{Huang2006} work shows that protein binding and lipid asymmetry, respectively, lead to precisely these kinds of dimpled domains.  

\begin{figure}
\begin{center}
\includegraphics[width=3.4in]{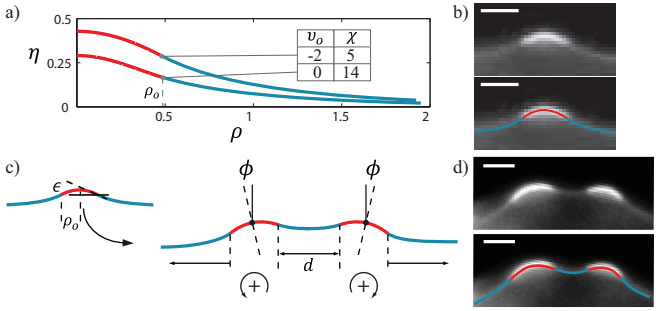}
\caption{\footnotesize Theoretical and experimental dimpled domain shapes.  Domains are shown in red, surrounding membrane in blue. a) Minimum energy dimples with and without spontaneous curvature (see legend, $\alpha=\pi/4$, $\sigma=1$).  b) Epi-fluorescence cross-section of a dimple on the surface of a GUV; the red and blue lines are a guide to the eye.  c) 1D model of interaction - dimples maintain shape, but tilt ($\phi$) as a function of separation distance ($d$).  Dimples with the same sign of curvature repel, while dimples with opposite sign attract.  d) Epi-fluorescence cross-section of two dimpled domains interacting on the surface of a GUV.  Scale bars are $3\,\mu\mbox{m}$.}
\label{fig3}
\end{center}
\end{figure}

Calculated shapes of dimpled domains induced by line tension and spontaneous curvature are shown in Figure \ref{fig3}a, alongside dimpled domains observed on giant unilamellar vesicles, shown in Figure \ref{fig3}b and d.\\

\n
{\bf Elastic Interactions of Dimpled Domains\\} 
Given two domains that have met the criteria for dimpling, the deformation in the membrane surrounding the domains mediates an elastic interaction when they are within a few elastic decay lengths ($\lambda_2$) of each other. This equips us to begin addressing how short length-scale and long time-scale membrane heterogeneity might be achieved. As previously stated, free diffusion sets the maximum rate at which a quenched membrane can evolve into a fully phase-separated membrane \cite{Bray2002}, where this evolution can happen in as little as a minute on the surface of a cell-sized vesicle \cite{Veatch2003}.  On the other hand, recycling and hence homogenization of cellular membrane is a process that takes place on the time-scale of an hour or more \cite{Hansen1992}.  Our measurements of domain interactions (detailed below and other data shown in SI) estimate the coalescence barrier between dimpled domains at $\sim5-10\,k_BT$.  Hence, given the diffusion-limited rate of phase separation, interactions slow this process by approximately $e^{-5}\simeq0.007$ to $e^{-10}\simeq0.00005$.  This makes the time-scale of lipid heterogeneity comparable to the time-scale of membrane recycling and even eukaryotic cell division.
 
The physical origin of domain interaction is explained by a simple model based on the assumption that the dimpled domain {\it shape} is constant during interaction, but the domains are free to tilt by an angle $\phi$, as shown in Figure \ref{fig3}c. This assumption was, in part, inspired by experimental observations of domain shapes on the surface of giant unilamellar vesicles, as shown, for example, in Figure \ref{fig3}d. The interaction energy is roughly an order of magnitude less than the free energy associated with the morphological transition itself (see Figure \ref{fig2}b), thus interaction does not perturb the domain shape significantly.  Only allowing domains to rotate simplifies the interaction between two domains to a change in the boundary conditions in the three regions of interest, shown in blue in Figure \ref{fig3}c.  Applying the small gradient approximation, the boundary slope is given by $|\epsilon-\phi|$ in the outer regions and by $|\epsilon+\phi|$ in the inner region.  With the single domain boundary slope, $\epsilon$, set by the energy minimization of the previous section ({\it i.e.} eqn.~\ref{eqsys}), the pairwise energy is minimized at every domain spacing, $d$, by $\partial G/\partial \phi = 0$ to find the domain tilt angle that minimizes the deformation energy (see SI for details).  This results in two qualitatively distinct scenarios:  two domains whose curvatures have the same sign repel each other, while two domains whose curvatures have the opposite sign attract each other.  Scaling arguments can be used to show that the strength of interaction between two dimpled domains increases roughly linearly with their area, so long as they are both larger than some critical area (see SI for details).  Mathematically, the assumption of rigidly rotating dimpled domains on a membrane is identical to a previous 2D model of bending-mediated interactions between intramembrane proteins represented by rigid conical inclusions \cite{Weikl1998}.  

Independent of the effects of spontaneous curvature, slight osmolar imbalances and constriction due to the lipid phase boundaries create small pressure gradients across the membrane that tend to orient all dimples in a cell or vesicle in the same direction, resulting in net repulsive interactions between all domains.  Transitions between `upward' and `downward' dimples are infrequent, due to a large energy barrier.  In the simplest case, where the domains are the same size, the tilt angle $\phi$ monotonically increases as two domains get closer, $\phi(d)\simeq-\epsilon e^{-d}$.  Likewise, the interaction energy, $V_{\mbox{\tiny int}}(d)\simeq2\pi\kappa_b\epsilon^2\rho_o^2e^{-d}$, increases monotonically with decreasing separation.  For direct comparison, we fit both the 1D model outlined here and the 2D inclusion model \cite{Weikl1998} to the data of Fig.\ref{fig4}, showing that they are experimentally indistinguishable, though with a slightly different elastic decay length. 

To quantitatively compare our interaction model with experiment, we examined the thermal motion of small domains on the surface of giant unilamellar vesicles, as described in `Materials and Methods.'  Membrane tension was regulated by balancing the internal and external osmolarity, giving us coarse control over the elastic decay length $\lambda_2$.  Through time, the distance between every domain pair was measured and the net results were used to construct a histogram.  The potential of mean force as a function of distance between domains is shown in Figure \ref{fig4}b.  We selected vesicles that had a low density of approximately equal-sized domains, and thus generally the interactions were described by a repulsive pairwise potential.  Though areal density of domains and generic data quality varied in our experiments (see SI), all data sets exhibit the repulsive core of the elastic interaction.  Multi-body interactions occur, though infrequently; their effect can be seen as a small variation in the baseline of Figure \ref{fig4}b, which is not captured by the pairwise interaction model.  At high membrane tension, when we would {\it not} expect dimpled domains, we qualitatively verified that domains coalesce in a rapid manner as compared to our low tension experiments (data not shown).  Other recent experiments have also observed repulsive interactions between domains on low membrane tension vesicles and the lack thereof on taut vesicles \cite{Yanagisawa2007}.

\begin{figure}
\begin{center}
\includegraphics[width=3.3in]{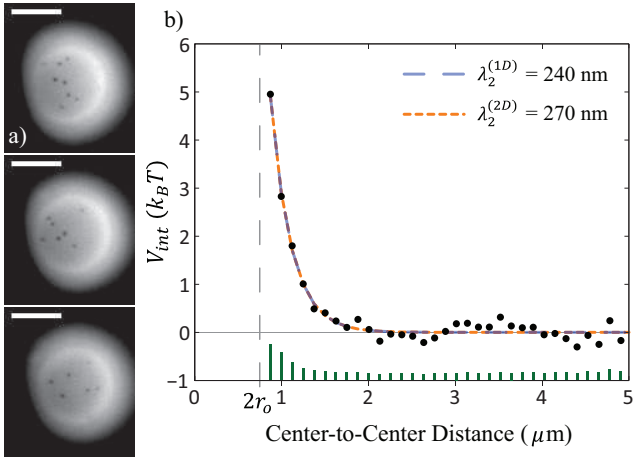}
\caption{\footnotesize Measuring domain interactions on the surface of a vesicle.  a)  Three images of dilute interacting domains on the surface of the same vesicle (scale bar is $10\,\mu\mbox{m}$).  b)  The repulsive interaction potential of domains on the surface of the same vesicle as (a).  The energy is measured in $k_BT$ and distance is domain center-to-center. The blue dashed line is a fit to the 1D interaction model in this paper, $V_{\mbox{\tiny int}}(r)= a_1e^{-r/\lambda_2^{\mbox{\tiny (1D)}}}+a_2$, with elastic decay length $\lambda_2^{\mbox{\tiny (1D)}}=240\,\mbox{nm}$.  The orange dashed line is a fit to the model, $V_{\mbox{\tiny int}}(r)= 2\pi\kappa_b\left[(a_1a_2)^2K_0(r/\lambda_2^{\mbox{\tiny (2D)}})+a_2^2a_3^4K_2^2(r/\lambda_2^{\mbox{\tiny (2D)}})\right]+a_4$, with elastic decay length $\lambda_2^{\mbox{\tiny (2D)}}=270\,\mbox{nm}$, based on the theory of Weikl {\it et al} \cite{Weikl1998}. Both elastic decay lengths indicate a membrane tension of $\sim4\times10^{-4}\,k_BT/\mbox{nm}^2$.  Errors bars are shown in green on the $x$-axis.}
\label{fig4}
\end{center}
\end{figure}

Our measurement of the pairwise potential allows us to estimate elastic properties of the membrane.  The elastic decay length was fit to the 1D and 2D interactions models described above, and found to be $\lambda_2^{\mbox{\tiny (1D)}}\simeq240\,\mbox{nm}$ and $\lambda_2^{\mbox{\tiny (2D)}}\simeq270\,\mbox{nm}$, respectively.  Taken with a nominal bending modulus of $25\,k_BT$, we estimate the membrane tension to be $\sim4\times10^{-4}\,k_BT/\mbox{nm}^2$.  From the images, we measure the size of the domains at $r_o\simeq350-400\,\mbox{nm}$, and hence $\rho_o\simeq1.5$.  We estimate the line tension, $\gamma$, using eqn.~\ref{chicrit}, based on the fact that the domains {\it are} dimpled, and find a lower bound of $\gamma\simeq0.49\,k_BT/\mbox{nm}$.  This is in good agreement with theoretical estimates and values determined from AFM measurements \cite{Schwille2007}, though somewhat higher than the value of $\gamma\simeq0.22\,k_BT/\mbox{nm}$ measured via shape analysis of fully phase separated vesicles \cite{Baumgart2003} and $\gamma\simeq0.40\,k_BT/\mbox{nm}$ from micropipette aspiration experiments \cite{Tian2007}.  Finally, viewing the repulsive core of the interaction as an effective activation barrier to coalescence, a simple Arrhenius argument suggests a decrease in coalescence kinetics by two to three orders of magnitude.  Indeed, such a slowing of coalescence was recently observed in a similar model membrane system \cite{Yanagisawa2007}.

\n\\
{\bf Discussion\\}
Comparing biologically relevant domain sizes ($\sim50-500\,\mbox{nm}$) with the elastic decay length ($\lambda_2$), we expect physiologically relevant domains to be small ({\it i.e.} small $\alpha$), as presumed in eqn.~\ref{chicrit}.  Estimating the elastic decay length requires knowledge of the membrane tension and bending stiffness.  We note that {\it in vitro} experiments of osmotically balanced single and multicomponent vesicles, and measurements of the plasma membrane of unstressed cells suggest membrane tensions of $10^{-4}-10^{-2}\,k_BT/\mbox{nm}^2$ \cite{Baumgart2003,Evans2000,Morris2001,Popescu2006}.  The typical bending modulus of a phosphocholine bilayer is $\sim10-50\,k_BT$, depending on the exact lipid and cholesterol content \cite{Evans2000,Chen1997,Ipsen2004}.  Choosing a nominal membrane tension of $10^{-4}\,k_BT/\mbox{nm}^2$ and nominal bending modulus of $25\,k_BT$ \cite{Evans2000,Baumgart2003} corresponds to an elastic decay length of $\lambda_2\simeq500\,\mbox{nm}$, suggesting that for lipid domains on the order of $50-500\,\mbox{nm}$, small $\alpha$ is an appropriate approximation.

Our experiments on the surface of GUVs have three potentially confounding effects, all due to the spherical curvature of the vesicle.  First, the surface metric is not entirely flat with respect to the image plane.  Thus, measurements of distance are underestimated the farther they are made from the projected vesicle center.  This problem is ameliorated by concentrating on domains which are at the bottom (or top) of the vesicle where the surface is nearly flat and demanding that our tracking software exclude domains that are out of focus; see SI for a more detailed explanation.  The second potential complication is that we use a flat 2D coordinate system for our theoretical analysis, however domains reside on a curved surface.  Given that the domain deformation, and hence energy density, decays exponentially with $\lambda_2$, as long as $\lambda_2$ is small with respect to the vesicle radius, the energetics that govern morphology converge on an essentially flat surface metric. The final complication is that the circular area of focus creates a fictitious confining potential for the domains, such that the effective measured potential of mean force is the sum of the elastic pairwise potential and a fictitious potential, $V_{\mbox{\tiny eff}}=V_{\mbox{\tiny int}}+V_{\mbox{\tiny fict}}$.  The fictitious potential is removed by simulating non-interacting particles in a circle the same size as the radius of focus (see SI for details).

The constant tension ensemble used in our theoretical analysis has a range of validity, determined by the excess area available on a thermally fluctuating membrane with conserved volume and total surface area $\mathcal{A}_o$ ({\it i.e.}~a vesicle).  In the limit where the morphological transitions use only a small portion ($\Delta\mathcal{A}$) of this excess area, defined by $k_BT/8\pi\kappa_b\gg\Delta\mathcal{A}/\mathcal{A}_o$, the tension is constant.  Outside this regime the tension rises exponentially with reduction in excess area, tending to stabilize dimples from fully budding (see SI for details).

In addition to the elastic mechanism of interaction, described herein, there may be other organizing forces at work in a phase-separated membrane: those of elastic \cite{Kuzmin2005}, entropic \cite{DeanPRE2006,Goulian1993} and electrostatic origin \cite{Groves2005}, however their putative length-scale, on the order of ten nanometers or less, is not accessible to the spatial and temporal resolution of our experiments, and not consistent with our measurement of an interaction length-scale of hundreds of nanometers.

\n\\
{\bf Conclusion\\}  
We have shown that lipid domains are subject to a morphological dimpling transition that depends on the bilayer elastic properties and domain size.  Dimpling allows two domains in proximity to repulsively interact due to the deformation in the surrounding membrane. Our model makes some key predictions:  at zero spontaneous curvature the domain size distribution reaches a critical point where coalescence is arrested by repulsive interactions; domains with finite spontaneous curvature are always subject to interaction and hence should always coalesce at a rate slower than the diffusion-limited rate. Additionally, the strength of elastic interactions is augmented by increasing line tension or domain area, with an approximately linear scaling.  The domain size and bilayer elastic parameters necessary to induce the dimpled morphology are consistent with physiological conditions.  Further, careful regulation of membrane cholesterol in cells may be related to the membrane mechanical properties necessary for morphological transitions. Combined with lipid recycling, our work offers a mechanism working against diffusion-driven coalescence, to maintain fine-scale lateral heterogeneity of lipids over time.

We proposed a simple 1D model of an elastic interaction that mediates dimpled-domain repulsion, and then used a standard ternary membrane system to verify the existence of dimpled domains and their subsequent repulsive interaction.  Finally, it follows that the morphologies and elastic forces which organize lipid domains might play an important role in the binding and lateral organization of proteins in the membrane.\\

\n
{\bf Materials and Methods\\} 
Giant unilamellar vesicles (GUVs) were prepared from a mixture of DOPC (1,2-Dioleoyl-sn-Glycero-3-Phosphocholine), DPPC (1,2-Dipalmitoyl-sn-Glycero-3-Phosphocholine) and cholesterol (Avanti Polar Lipids) (25:55:20/molar) that exhibits liquid-liquid phase coexistence \cite{Veatch2003}.  Fluorescence contrast between the two lipid phases is provided by the rhodamine head-group labeled lipids: DOPE (1,2-Dioleoyl-sn-Glycero-3-Phosphoethanolamine-N- (Lissamine Rhodamine B Sulfonyl)) or DPPE (1,2-Dipalmitoyl-sn-Glycero-3-Phosphoethanolamine-N- (Lissamine Rhodamine B Sulfonyl)), at a molar fraction of $\sim0.005$.  The leaflet compositions are presumed symmetric and hence $\upsilon_o=0$.  

GUVs were formed via electroformation \cite{Veatch2003,Angelova1992}.  Briefly, $3-4\,\mu\mbox{g}$ of lipid in chloroform were deposited on an indium-tin oxide coated slide and dessicated for $\sim2\,\mbox{hrs}$ to remove excess solvent.  The film was then hydrated with a $100\,\mbox{mM}$ sucrose solution and heated to $\sim50\,\mbox{C}$ to be above the miscibility transition temperature.  An alternating electric field was applied; $10\,\mbox{Hz}$ for 120 minutes, $2\,\mbox{Hz}$ for 50 minutes, at $\sim500\,\mbox{Volts/m}$ over $\sim2\,\mbox{mm}$.  Low membrane tensions were achieved by careful osmolar balancing with sucrose ($\sim100\,\mbox{mM}$) inside the vesicles, and glucose ($\sim100-108\,\mbox{mM}$) outside. 

Domains were induced by a temperature quench (see SI) and imaged using standard TRITC epi-fluorescence microscopy at 80x magnification with a cooled (-30 C) CCD camera (Roper Scientific, $6.7\times6.7\,\mu\mbox{m}^2$ per pixel, 20 MHz digitization).  Images were taken from the top or bottom of a GUV where the surface metric is approximately flat (see SI).  Data sets contained $\sim500-1500$ frames collected at 10-20 Hz with a varying number of domains (usually $5-10$).  The frame rate was chosen to minimize exposure-time blurring of the domains, while allowing sufficiently large diffusive domain motion.   Software was written to track the position of each well-resolved domain and calculate the radial distribution function.  The raw radial distribution function was corrected for the fictitious confining potential of the circular geometry (see SI).  In the dilute interaction limit, pairwise interactions dominate, and the negative natural logarithm of the radial distribution function is the interaction potential (potential of mean force) plus a constant, as shown in Figure \ref{fig4}b.\\

\footnotesize{We thank Patricia Bassereau, Ben Freund, Kerwyn Huang, Greg Huber, Sarah Keller and Udo Seifert for stimulating discussion and comments on the manuscript, and Jenny Hsaio for help with experiments.  TU and RP acknowledge the support of the National Science Foundation award No.~CMS-0301657, NSF CIMMS award No.~ACI-0204932, NIRT award No.~CMS-0404031 and the National Institutes of Health Director's Pioneer Award. WK acknowledges support from NSF CAREER Award CMMI-0748034.}


}
{\tiny

}
\end{document}